\documentclass[aps,pra,reprint,footinbib,superscriptaddress,longbibliography]{revtex4-2}

\usepackage[utf8]{inputenc}  
\usepackage[T1]{fontenc}     
\usepackage[american]{babel}  
\usepackage[svgnames,psnames]{xcolor}
\usepackage{hyperref}
\usepackage{graphicx}
\hypersetup{colorlinks=true, citecolor=FireBrick, urlcolor=blue, linkcolor=blue}
\usepackage{amsmath,amssymb,amsthm,bm,mathtools,amsfonts,mathrsfs,bbm,dsfont,times,appendix,multirow,color,cases} 
\usepackage{lipsum}
\usepackage{natbib}
\usepackage[ruled]{algorithm2e}
\usepackage{verbatim}
\usepackage{multirow}

\newcommand{\tr}{{\mathrm{tr}}}

\newcommand{\eins}{\mathbbm{1}}

\renewcommand{\vr}{\ensuremath{\varrho}}

\usepackage{braket}
\usepackage[normalem]{ulem}
\usepackage[ruled]{algorithm2e}

\begin{document}
\title{Variational-toolbox-based separability detection of multiqubit states}
\author{Jin-Min Liang}
\email{jmliang@pku.edu.cn}
\affiliation{State Key Laboratory for Mesoscopic Physics, School of Physics, Frontiers Science Center for Nano-optoelectronics, $\&$ Collaborative Innovation Center of Quantum Matter, Peking University, Beijing 100871, China}
\author{Shao-Ming Fei}
\email{feishm@cnu.edu.cn}
\affiliation{School of Mathematical Sciences, Capital Normal University, Beijing 100048, China}
\author{Qiongyi He}
\email{qiongyihe@pku.edu.cn}
\affiliation{State Key Laboratory for Mesoscopic Physics, School of Physics, Frontiers Science Center for Nano-optoelectronics, $\&$ Collaborative Innovation Center of Quantum Matter, Peking University, Beijing 100871, China}
\affiliation{Collaborative Innovation Center of Extreme Optics, Shanxi University, Taiyuan, Shanxi 030006, China}
\affiliation{Hefei National Laboratory, Hefei 230088, China}
\date{\today}

\begin{abstract}
    Parameterized quantum circuits (PQCs) play a crucial role in variational quantum algorithms. While it is commonly believed that the optimal PQC can be used to reproduce the target state, here we reveal that the optimal PQC can also provide valuable insights into the state's properties. We propose variational toolboxes to identify the $k$-separability of pure states, with or without preparation noise, by checking the structure within the optimal PQCs. Additionally, we introduce adaptive optimization strategies to detect the $k$-separability of mixed states. Compared to fixed PQCs, our approach controls fewer parameters for low-rank states. Finally, we validate our methods through numerical demonstrations for various states.
\end{abstract}

\maketitle

\section{Introduction}
Entanglement is one of the important resources in quantum-based applications, such as quantum communication~\cite{hu2021long,bulla2023nonlocal}, quantum network~\cite{pompili2021realization}, and quantum metrology~\cite{colombo2022time,long2022entanglement}. Understanding the entanglement property of a state enables us to efficiently manipulate quantum states in practical situations. One of the essential issues is the detection of entanglement for pure or mixed states, which confirms whether a target state is entangled. Another crucial problem is determining the entanglement structure of an entangled state~\cite{guhne2009entanglement,shahandeh2014structural,lu2018entanglement}. In principle, multiparticle quantum states may be entangled as a whole system, called genuine multipartite entanglement (GME), or entangled only involving different subsystems. The entanglement structure of a quantum state is mathematically described by the $k$-separability or $m$-producibility~\cite{sorensen2001entanglement}. Both questions are challenging because of exponentially large systems. Thus, exploring efficient methods to identify the presence of entanglement and the detailed structure is a crucial theoretical and experimental problem.
    
Recent decades have seen significant advances in entanglement detection methods, including the positive partial transpose criterion~\cite{horodecki1997separability}, the computable cross norm criterion~\cite{chen2003matrix,rudolph2003some}, the permutation criterion~\cite{horodecki2006separability,liu2022detecting}, and machine learning methods~\cite{girardin2022building,chen2023certifying}. These methods highly depend on the prior information of the state or the labeled dataset to train the classical neural network. As the system size increases, quantum state tomography~\cite{cramer2010efficient} and the determination of the training state become infeasible. On the other hand, extracting the partial information of quantum states may also characterize the entanglement by employing randomized measurements~\cite{elben2020mixed,neven2021symmetry,yu2021optimal,ketterer2022statistically,cieslinski2024analysing}.
    
The advent of quantum techniques provides potential opportunities to characterize the quantum correlation properties of multiqubit states prepared on noisy intermediate-scale quantum devices (NISQs)~\cite{preskill2018quantum}. Well-known quantum techniques include variational quantum algorithms (VQAs), parameterized quantum circuits (PQCs), the SWAP and HADAMARD test, etc.~\cite{cerezo2021variational,bharti2022noisy}. These techniques have been adopted as useful tools for quantum simulation~\cite{cervera2021meta}, machine learning~\cite{chen2021universal}, and the characterization of quantum entanglement~\cite{bravo2020quantum,beckey2021computable,moller2022variational,wang2022detecting,chen2023near,liu2023solving,consiglio2022variational,philip2024schrodinger}.

Current VQAs focus on constructing PQCs and learning an optimal PQC to encode specific states by a hybrid quantum-classical optimization process~\cite{cerezo2021variational,bharti2022noisy}. Following the training stage, the properties of the encoded state can be extracted using postprocessing methods~\cite{huang2020predicting}. Traditionally, state preparation and analysis are treated as separate components, with little consideration for interconnection. Since the target state is prepared by implementing the optimal PQC on an initial state, the optimal PQC inherently encapsulates all information about the target state. However, limited research has delved into the relationship between the optimal PQC and the extraction of information from the reproduced state.
    
Focusing on one of the important properties, the separability of multiqubit states~\cite{nagata2002configuration,dur2000classification,seevinck2008partial}, recent work only considered the relation between the fully separable states and the structure of the optimal unitary~\cite{consiglio2022variational}. On the one hand, multiqubit states exhibit sophisticated entanglement structures, such as partial separable ($k$ separable), which has been identified as a resource in quantum metrology~\cite{gessner2018sensitivity,ren2021metrological}. On the other hand, noise is unavoidable when preparing a multiqubit pure state. Thus, applying known methods to experimentally prepared states could lead to incorrect results regarding separability. Finally, the number of parameters within the PQC increases exponentially with the number of qubits for mixed states, which is extremely challenging for a larger number of qubits. These issues motivate us to develop efficient strategies to verify the separability of quantum states.
    
In this work, we address the above issues and utilize the framework of VQAs to explore the connection between the structure of the optimal PQC and the separability of a multiqubit state. First, we construct two different PQC pools, and each PQC in the pools can be used to reconstruct a target state. Next, we propose a method to detect the separability of pure multiqubit states and generalize the methods to the noise situation by introducing virtual distillation~\cite{huggins2021virtual,koczor2021exponential}. Moreover, we propose two adaptive quantum algorithms for detecting the separability of mixed states. The parameterized quantum state is a probability ensemble of parameterized quantum pure states generated via the PQC from the pools. Finally, numerical results are presented to verify our approaches. The contributions of our results are twofold. On the one hand, we can not only detect the $k$-separability of a state but also confirm the entanglement structure between qubits. On the other hand, our approach is more efficient for low-rank mixed states.
    
\section{The parameterized quantum circuit pools}
An $n$-qubit pure state $|\psi\rangle$ is a $k$-separable state if it can be written as a tensor product of $k$ pure states, $|\psi\rangle=\otimes_{i=1}^{k}|\psi_i\rangle$, and a mixed state is $k$-separable if it is a convex mixture of $k$-separable pure states~\cite{nagata2002configuration,dur2000classification,seevinck2008partial}. To detect the $k$-separability of quantum states, we construct a family of PQCs. The primitive unitaries are two single-qubit unitaries
\begin{align}
    R_z(\alpha)=e^{\iota\alpha Z/2},~R_y(\beta)=e^{\iota\beta Y/2},
\end{align}
and a two-qubit interaction unitary~\cite{kraus2001optimal}
\begin{align}\label{II:eq1}
    Q(\boldsymbol{\theta})\!&\!=\!\mathrm{C}_{2,1}\!\left[ R_z\left(\theta_1\right)\otimes R_y\left(\theta_2\right)\right]\!
    \mathrm{C}_{1,2}\!\left[\mathds{1}\otimes R_y\left(\theta_3\right)\right]\!\mathrm{C}_{2,1},
\end{align}
where $\iota$ denotes the imaginary unit, $\alpha,\beta,\theta_1$,$\theta_2$,$\theta_3$ are parameters, and the unitary $\mathrm{C}_{i,j}$ means that for the controlled-not gate, the controlled qubit is $i$ and the work qubit is $j$. See details in Appendix~\ref{AppendixA}.
    
Based on primitive unitaries, we construct a local unitary
\begin{align}\label{II:eq2}
    W(\boldsymbol{\alpha})=
    \bigotimes_{i=1}^{n}R_z(\alpha_{3,i})R_y(\alpha_{2,i})R_z(\alpha_{1,i}),
\end{align}
and $L$ unitaries $\{V_l(\boldsymbol{\alpha},\boldsymbol{\theta})\}_{l=1}^{L}$, where $L$ is an integer depending on $n$. Each unitary
\begin{align}\label{II:eq3}
    V_l(\boldsymbol{\alpha},\boldsymbol{\theta})=\!W(\boldsymbol{\alpha})\mathcal{Q}_l(\boldsymbol{\theta}),
\end{align}
and $\mathcal{Q}_l(\boldsymbol{\theta})$ is a product of two-qubit interaction unitary $Q_{j,k}(\boldsymbol{\theta})$ acting on the $j$th and $k$th qubits, $j,k\in\{1,\cdots,n\}$, $j\neq k$.
    
For an $n$-qubit system, there are $\binom{n}{2}$ pairs of $(j,k)$ and therefore $\binom{n}{2}$ different unitaries $Q_{j,k}(\boldsymbol{\theta})$. However, qubit pairs $(i_1,i_2)$ and $(i_3,i_4)$ can be processed simultaneously for different qubits $i_1\neq i_2\neq i_3\neq i_4$. With a parallelized way, only $L=\frac{3n-2}{2}$ $\left(\frac{3n-3}{2}\right)$ unitaries for even (odd) $n$ are required to run over all $Q_{j,k}(\boldsymbol{\theta})$. For even $n$, we write the first two unitaries,
\begin{align}
    V_1(\boldsymbol{\alpha},\boldsymbol{\theta})
    \!&\!=\!W(\boldsymbol{\alpha})\left[\bigotimes_{i=1}^{n-1}Q_{i,i+1}(\boldsymbol{\theta}_{i})\right],\label{II:eq4}\\
    V_2(\boldsymbol{\alpha},\boldsymbol{\theta})
    \!&\!=\!W(\boldsymbol{\alpha})\left\{Q_{1,n}(\boldsymbol{\theta}_{1})\otimes\left(\bigotimes_{j=2}^{n-2}Q_{j,j+1}(\boldsymbol{\theta}_{j})\right)\right\}\label{II:eq5},
\end{align}
where $i$ runs over odd numbers and $j$ runs over even numbers. Detailed PQCs for $4$- and $10$-qubit systems are presented in Appendix~\ref{AppendixA}. We remark that if we aim to verify the full separability of a multiqubit state, the local unitary $W(\boldsymbol{\alpha})$ shown in Eq.(\ref{II:eq2}) can be reduced into $W(\boldsymbol{\alpha})=\bigotimes_{i=1}^{n}R_z(\alpha_{2,i})R_y(\alpha_{1,i})$ containing only two single-qubit unitaries~\cite{nielsen2010quantum}.

Combining the structure of PQCs $W$ and $V_l$ in Eqs.(\ref{II:eq2}) and (\ref{II:eq3}), we construct two PQC pools
\begin{align}
    \mathcal{P}_1&=\left\{W(\boldsymbol{\alpha})\right\},\label{II:eq6}\\
    \mathcal{P}_2&=\left\{V_l\left(\boldsymbol{\alpha},\boldsymbol{\theta}\right)W\left(\boldsymbol{\gamma}\right)\right\}_{l=1}^{L}.\label{II:eq7}
\end{align}
The parameters $\boldsymbol{\alpha},\boldsymbol{\theta},\boldsymbol{\gamma}$ are tunable.
The total number of PQCs in $\mathcal{P}_1$ and $\mathcal{P}_2$ are respectively $|\mathcal{P}_1|=1$ and $|\mathcal{P}_2|=L$.

\section{Multiqubit pure state separability detection}
In this section, we focus on the problem of determining the separability of multiqubit pure states. In Sec. \ref{SecIII:A}, we propose methods for identifying the separability of ideal pure states that are perfectly prepared. However, in practical situations, noise in the NISQ device leads to the production of mixed states instead of pure states. In Sec. \ref{SecIII:B}, we consider a noise situation in which the obtained state is a noisy mixed state.

\subsection{Ideal multiqubit pure state}\label{SecIII:A}
The first step to detect the separability of an $n$-qubit pure state $|\psi\rangle$ is a variational quantum state reconstruction (VQSR), where the goal is to reconstruct the quantum state $|\psi\rangle$ using a VQA.
    
The PQC are chosen from two PQC pools, $\mathcal{P}_1$ and $\mathcal{P}_2$ corresponding to Eqs.(\ref{II:eq6}) and (\ref{II:eq7}). We optimize the cost function
\begin{align}\label{III:eq1}
    F(\boldsymbol{\Phi})=1-\sqrt{\tr[O_GU^{\dag}(\boldsymbol{\Phi})|\psi\rangle\langle\psi|U(\boldsymbol{\Phi})]},
\end{align}
where $O_G=|0^{\otimes n}\rangle\langle0^{\otimes n}|$. Eq.(\ref{III:eq1}) can be expressed in terms of the fidelity between the state $U(\boldsymbol{\Phi})|0^{\otimes n}\rangle$ and the target state $|\psi\rangle$,
\begin{align}\label{III:eq2}
    F(\boldsymbol{\Phi})=1-\sqrt{|\langle0^{\otimes n}|U^{\dag}(\boldsymbol{\Phi})|\psi\rangle|^2}.
\end{align}
When $F(\boldsymbol{\Phi}^{*})=0$, then $|\psi\rangle=U(\boldsymbol{\Phi}^{*})|0^{\otimes n}\rangle$, where $\boldsymbol{\Phi}^{*}$ represents the optimal parameter. During the optimization process, the cost function $F(\boldsymbol{\Phi})$ and its gradients can be evaluated on a near-term quantum computer using a shallow-depth circuit and a single copy of the state $|\psi\rangle$~\cite{ekert2002direct,quek2024multivariate}. The Hoeffding bound~\cite{hoeffding1994probability} suggests that the measurements $O(\varepsilon^{-2}\log_2\delta^{-1})$ are sufficient to evaluate $F(\boldsymbol{\Phi})$ with an additive error of $\varepsilon$ and a success probability of at least $1-\delta$~\cite{buhrman2001quantum}.
    
The training process involves selecting different PQCs from sets $\mathcal{P}_1$ and $\mathcal{P}_2$ as the unitary $U(\boldsymbol{\Phi})$. We iterate through $l=1,2$ and use VQSR to find the global minimum of $F(\boldsymbol{\Phi})$ by choosing a PQC from $\mathcal{P}_l$. The training procedure is terminated if the function value is below a threshold $\epsilon$ ($0<\epsilon\ll1$). If the minimum value does not meet the termination condition for any unitary in sets $\mathcal{P}_1$ and $\mathcal{P}_2$, then our approach is unable to determine the separability of $|\psi\rangle$.
    
Once the optimal parameters are found, we proceed to the postprocessing step, using the obtained optimal parameters and the corresponding structure of the parameterized quantum circuit to infer the separability of the pure state $|\phi\rangle$. The analysis is divided into two cases.
 
\textit{Case 1.} If the optimal PQC comes from $\mathcal{P}_1$, it indicates that the state $|\psi\rangle$ is fully separable. This is clear because the PQC in $\mathcal{P}_1$ is a local unitary that does not create any entanglement.
    
\begin{figure}[ht]
    \centering
    \includegraphics[scale=0.37]{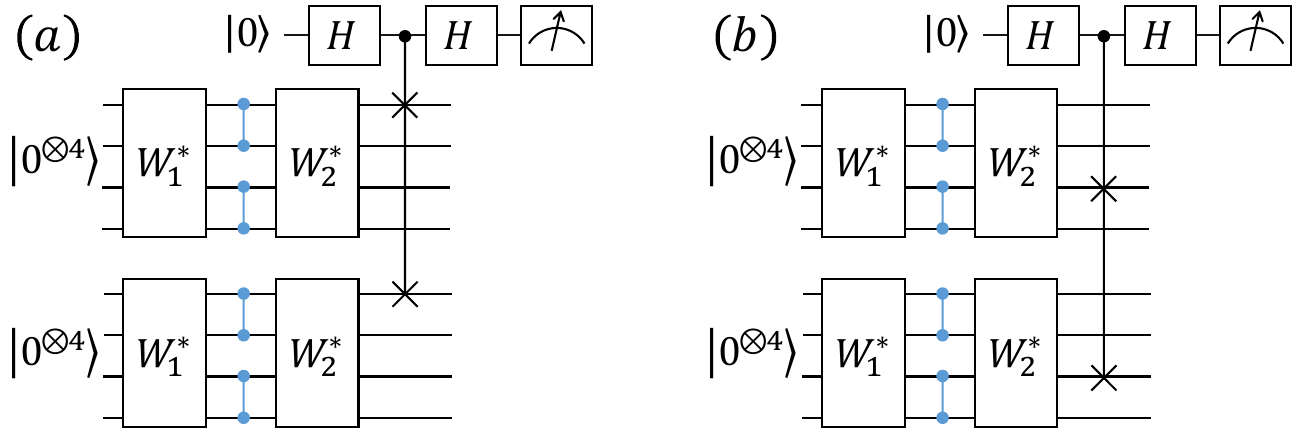}
    \caption{A $4$-qubit example. (a) The circuit for calculating the purity $\tr(\varrho_1^2)$. (b) The circuit for calculating the purity $\tr(\varrho_3^2)$.}
    \label{Fig1}
\end{figure}

\textit{Case 2.} If the optimal PQC comes from $\mathcal{P}_2$, we need to examine whether the two-qubit interaction unitary $Q_{j,k}(\boldsymbol{\theta}^{*})$ creates entanglement when applying it to a two-qubit separable state. We evaluate the entanglement property by calculating the purity $\tr(\varrho_j^2)$ of the reduced density matrix $\varrho_j=\tr_{\{1,\cdots,n\}/j}(|\psi\rangle\langle\psi|)$. Figure \ref{Fig1} shows an example of $4$-qubit system. If the purity is not one, it indicates that entanglement is formed between the qubits $(j,k)$. In this case, an edge is added between the qubit pair $(j,k)$. Conversely, if the purity is equal to one, it implies that the qubits $(j,k)$ remain separable, and no edge is added between them.

After examining all the two-qubit interaction unitaries, we construct a graph $G=(V,E)$ with $n$ vertices $V=\{1,2,\cdots,n\}$ representing individual qubits, and edges $E=\{E_{ij}\}$ for different $(i,j)$, where $E_{ij}$ denotes that the vertices (qubit pair) $(i,j)$ are connected. Based on the graph, we conclude that an $n$-qubit pure state is $k$-separable if the graph $G$ can be partitioned into at most $k$ disconnected subgraphs.
 
Two remarks are as follows. On the one hand, the defined PQCs in our framework have only one entangled layer. This feature implies our approach is only suitable for states that can be prepared by the defined PQCs. In particular, the two-qubit interaction between $n$ separable qubits generates at most $\lceil\frac{n}{2}\rceil$-separable states, where $\lceil\cdot\rceil$ is the ceiling function. For more general states involving three-or-more-qubit interactions, our approach fails to detect the separability. Although adding more entangled layers can enhance the expressibility of PQC, we fail to infer the separability from the learned PQC. On the other hand, if one is interested in the separability of an $n$-qubit pure state under a specific bipartition such as $A|B$, we can construct the PQC as $U=U_A\otimes U_B$ and then use our framework to detect its separability, where $U_A,U_B$ are Hardware-efficient Ansatz commonly used in VQA~\cite{kandala2017hardware,tand2021qubit}.
	
\subsection{Noisy multiqubit pure state}\label{SecIII:B}
Suppose we can detect the separability of the ideal $n$-qubit pure state $|\psi\rangle$ by the approach in Sec.~\ref{SecIII:A}. However, one experimentally obtains a noisy mixed state under some noise channel $\mathcal{E}$~\cite{nielsen2010quantum},
\begin{align}
    \varrho_{\mathrm{noise}}&=\mathcal{E}(|\psi\rangle\langle\psi|)\\
    &=(1-q)|\psi\rangle\langle\psi|+q\sum_{i}E_i|\psi\rangle\langle\psi| E_i^{\dag},
\end{align}
with $\sum_{i}E_i^{\dag}E_i=\mathds{1}_{2^n}$. In this scenario, inferring the $k$-separability of $|\psi\rangle$ from the obtained noisy states $\varrho_{\mathrm{noise}}$ is a challenge. If we directly utilize ordinary detection methods for the mixed state~\cite{ketterer2022statistically}, we may obtain a wrong conclusion. Next, we show that it is possible to infer the separability of $|\psi\rangle$ from the noisy mixed state.
    
Assume that the preparation error can be mitigated using the purification-based quantum error mitigation method~\cite{koczor2021exponential,huggins2021virtual} such that
\begin{align}
    \lim_{m\to+\infty}\frac{\varrho_{\mathrm{noise}}^{m}}{\tr[\varrho_{\mathrm{noise}}^m]}=|\psi\rangle\langle\psi|
\end{align}
where $m$ is an integer~\cite{endo2021hybrid,cai2023quantum}. To detect the $k$-separability of $|\psi\rangle$, a naive approach is to prepare the state $\frac{\varrho_{\mathrm{noise}}^{m}}{\tr[\varrho_{\mathrm{noise}}^m]}$ and perform ordinary detection approaches~\cite{ketterer2022statistically}. However, the preparation of $\frac{\varrho_{\mathrm{noise}}^{m}}{\tr[\varrho_{\mathrm{noise}}^m]}$ is a difficult problem~\cite{tan2021variational}. Following the approach for the ideal multiqubit pure state, we only need to define a new cost function
\begin{align}\label{EqNoise}
    F(\boldsymbol{\Phi})&=1-\sqrt{\frac{\tr[O_G\tilde{\varrho}^{m}]}{\tr[\varrho_{\mathrm{noise}}^m]}},
\end{align}
for an integer $m$ and the state $\tilde{\varrho}=U^{\dag}(\boldsymbol{\Phi})\varrho_{\mathrm{noise}}U(\boldsymbol{\Phi})$. The numerator and denominator can be estimated using shallow depth quantum circuits~\cite{quek2024multivariate} or classical shadow techniques~\cite{huang2020predicting,zhou2024hybrid}.

Consider the global depolarizing noise model~\cite{nielsen2010quantum}, $\varrho_{\mathrm{noise}}=(1-q)|\psi\rangle\langle\psi|+q\frac{\mathds{1}_{2^n}}{2^n}$, $q\in[0,1]$. Our approach can detect the separability of $|\psi\rangle$ for any noise strength $q$ since there always exists an integer $m$ such that $\frac{\varrho_{\mathrm{noise}}^{m}}{\tr[\varrho_{\mathrm{noise}}^m]}\approx|\psi\rangle\langle\psi|$ based on the power iteration~\cite{golub2013matrix}; for instance, the $10$-qubit mixed state,
\begin{align}\label{V:eq1}
    \varrho_{g}(q)&=(1-q)|\mbox{Bell}^{\otimes5}\rangle\langle\mbox{Bell}^{\otimes5}|+\frac{q}{2^{10}}\mathds{1}_{2^{10}},
\end{align}
where the noise strength $q\in[0,1]$ and each Bell state is $|\mbox{Bell}\rangle=(|01\rangle+|10\rangle)/\sqrt{2}$. It is clear that the source state $|\mbox{Bell}^{\otimes5}\rangle\langle\mbox{Bell}^{\otimes5}|$ is $5$-separable state. To detect the entanglement of the state $|\mbox{Bell}^{\otimes5}\rangle$ from the noisy mixed state $\varrho_{g}(q)$, we use $m$ copies of $\varrho_{g}(q)$ and learn the state $\frac{\varrho_{\mathrm{noise}}^{m}}{\tr[\varrho_{\mathrm{noise}}^m]}$ by the cost function~(\ref{EqNoise}). Figure \ref{Fig3}(a) shows the infidelity
\begin{align}\label{V:eq2}
    f(q)=\left|1-\frac{\langle\mbox{Bell}^{\otimes5}|\varrho_{g}^{m}(q)|\mbox{Bell}^{\otimes5}\rangle}{\tr[\varrho_{g}^m(q)]}\right|,
\end{align}
as a function of $m$. We find that it is possible to learn the entanglement of the state $|\mbox{Bell}^{\otimes5}\rangle$ by using $m\leq5$ copies of the mixed state $\varrho_{g}(q)$ for $q=0.2,0.4,0.6,0.8$. The greater the noise strength, the more copies of the noisy state are required.
 
\section{Mixed state separability detection with adaptive optimization}\label{SecIV}
Generalizing the pure state to a mixed $n$-qubit state $\varrho$ is feasible by finding a convex roof extension~\cite{bennett1996mixed,uhlmann2010roofs}. To solve the full separability of an $n$-qubit mixed state, one needs to construct a fully separable parameterized mixed state~\cite{consiglio2022variational},
\begin{align}\label{IV:eq1}
    \sigma=\sum_{m=1}^{M}q_m\sigma_m,
\end{align}
where $\sum_{m=1}^{M}q_m=1$ and each pure state $\sigma_m$ is a fully separable state. If there is a state $\sigma$ satisfying the Hilbert-Schmidt distance
\begin{align}\label{IV:eq2}
    F=\|\varrho-\sigma\|_{\mathrm{HS}}^2\leq\epsilon,
\end{align}
for a small threshold $\epsilon$, we confirm that $\varrho$ is fully separable. The number of parameters to generate the decomposition [Eq. (\ref{IV:eq1})] is $M(1+2n)$ and the calculation of $F$ requires $O(M^2)$ items. According to Carathéodory's theorem~\cite{horodecki1997separability,vedral1998entanglement}, $M\leq4^n$. One common approach sets $M=4^n$ to detect any state $\varrho$. This fixed parametrization strategy naturally suffers from the exponentially large classical parameters. It would be challenging for a large number of qubits. However, if the target state $\varrho$ is a low-rank state such as $R(\varrho)=\Omega(n)$, then $M=O(n^2)$ suffices to learning the state $\varrho$, where $R(\varrho)$ denotes the rank of $\vr$. In this scenario, the fixed parametrization method would be inefficient.
	
Here, we propose an adaptive method to detect whether a mixed state is fully separable. The algorithm is summarized as Algorithm 1.

\textbf{Algorithm 1}. Adaptive method for full separability detection of a mixed state.

Given a target mixed state $\varrho$ and a threshold $\epsilon$, perform the following steps for $S=1,2,\cdots$:

\textbf{Step 1}. Prepare an $n$-qubit fully separable state
    \begin{align}\label{IV:eq3}
        \sigma(\boldsymbol{\Phi})=\sum_{m=1}^{S}q_m\sigma_m(\boldsymbol{\alpha}_m)
    \end{align}
with $S(2n+1)$ parameters $\boldsymbol{\Phi}=\{(q_m,\boldsymbol{\alpha}_m)\}_{m=1}^{S}$.
  
\textbf{Step 2}. Minimize the cost function
\begin{align}\label{IV:eq4}
    F_{S}(\boldsymbol{\Phi})=\|\sigma(\boldsymbol{\Phi})-\varrho\|_{\rm{HS}}^2,
\end{align}
by the classical-quantum feedback loop and find optimal parameters $\boldsymbol{\Phi}^{*}$.
  
\textbf{Step 3}. If the value of the function $F_{S}(\boldsymbol{\Phi}^{*})$ is lower than the threshold $\epsilon$, we terminate the loop and conclude that $\varrho$ is an $n$-separable state. Otherwise, we obtain nothing about the entanglement of $\varrho$.
    
Based on Eq.(\ref{IV:eq3}), the number of parameters of Algorithm 1 for each optimization is $S(1+2n)$. As a comparison, the number of parameters of the fixed parametrization method~\cite{consiglio2022variational} is $4^n(1+2n)$. Thus, Algorithm 1 is efficient for states with rank $R(\vr)=\Omega(n)$ which ensures $S=1,\cdots,O(n^2)$.
    
Following the adaptive optimization approach of Algorithm 1, we propose a mixed state $k$-separability detection summarized as Algorithm 2.

\textbf{Algorithm 2}. Adaptive method for $k$-separability detection of a mixed state.

Given a target mixed state $\varrho$ and a threshold $\epsilon$, perform the following steps for $S=1,2,\cdots$.

\textbf{Step 1}. Prepare an $n$-qubit mixed state
\begin{align}\label{IV:eq5}
    \sigma(\boldsymbol{\Phi}_1)=\sum_{k=1}^{S}\sum_{l=1}^{L}q_{l}^{k}\sigma_{l}^{k},
\end{align}
with classical probability parameters $q=\{q_{l}^{k}\}$ and each pure state $\sigma_{l}^{k}=|\sigma_{l}\rangle\langle\sigma_{l}|$, where
\begin{align}\label{IV:eq6}
    |\sigma_{l}\rangle=
    V_l(\boldsymbol{\alpha}_{l}^{1},\boldsymbol{\theta}_{l}^{1})
    W(\boldsymbol{\alpha}_{l}^{0})|0^{\otimes n}\rangle
\end{align}
for $l=1,\cdots,L$ and $|\sigma_{0}\rangle=W(\boldsymbol{\alpha})|0^{\otimes n}\rangle$.
        
\textbf{Step 2}. Minimize the cost function
\begin{align}\label{IV:eq7}
    \hat{F}_{S}(\boldsymbol{\Phi}_{1})=
    \|\sigma(\boldsymbol{\Phi}_{1})-\varrho\|_{\rm{HS}}^2,
\end{align}
by the classical-quantum feedback loop and find optimal parameters $\boldsymbol{\Phi}_{1}^{*}$.
		
\textbf{Step 3}. If the value of the function $\hat{F}_{S}(\boldsymbol{\Phi}_{1}^{*})$ is lower than the threshold $\epsilon$, we terminate the loop and output a convex-roof decomposition of $\varrho$,
\begin{align}\label{IV:eq8}
    \sigma(\boldsymbol{\Phi}^{*})=\sum_mq_m^{*}\sigma_m^{*}.
\end{align}

\textbf{Step 4}. Use the postprocessing step in Sec.~\ref{SecIII:A} to detect the separability of each pure state $\sigma_{m}^{*}$ in Eq.(\ref{IV:eq8}). Suppose $\sigma_{m}^{*}$ is $k_{m}$-separable. Then $\varrho$ is $k$-separable, where $k=\min_mk_{m}$.
    
\begin{figure*}[ht]
    \centering
    \includegraphics[scale=0.37]{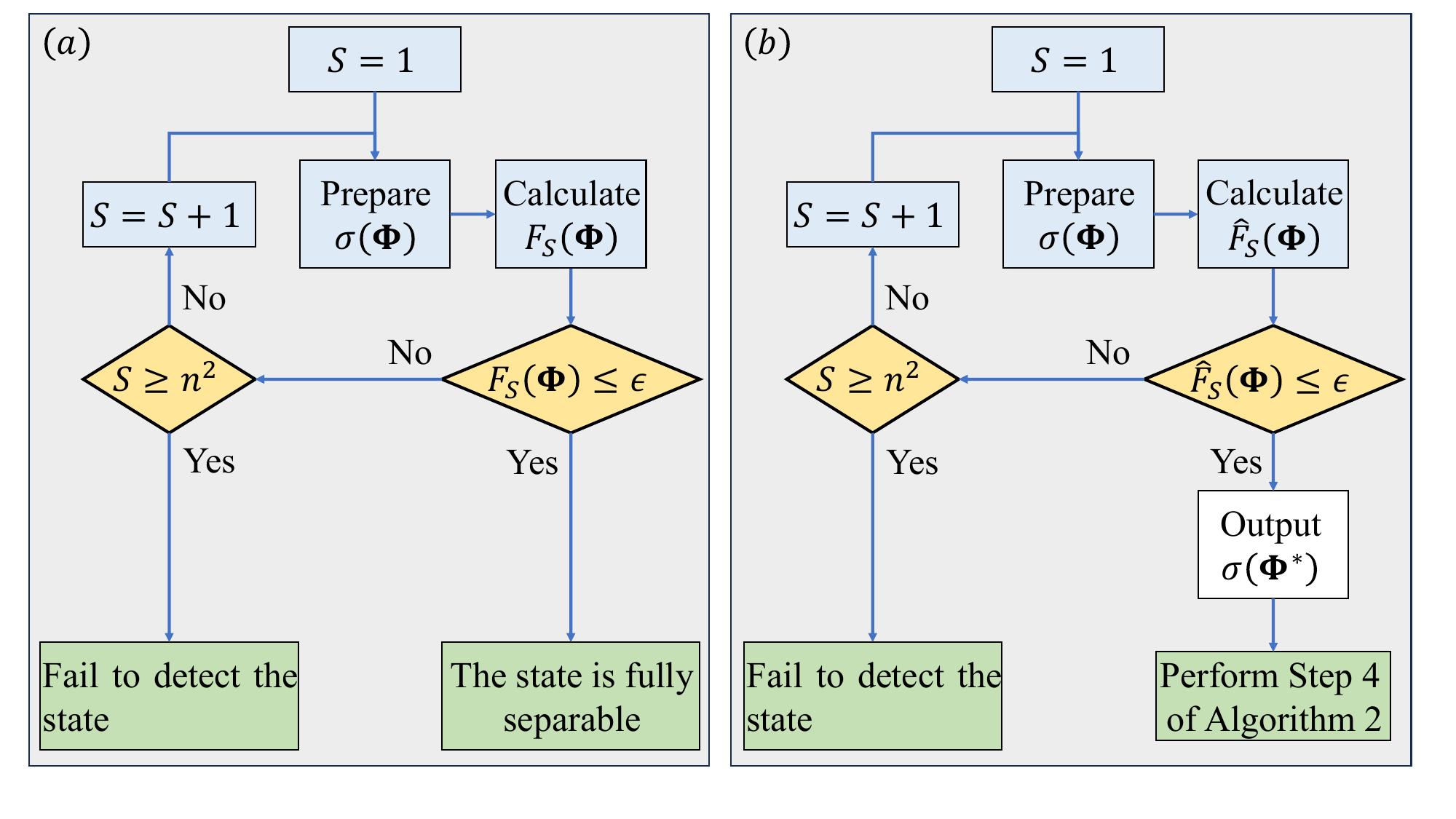}
    \caption{Schematic illustration of (a) algorithm 1 and (b) algorithm 2. Note that we set a threshold for $S$ as $n^2$, ensuring our algorithms are suitable for near-term quantum devices.}
    \label{Fig2}
\end{figure*}

Here, all PQCs in $\mathcal{P}_2$ are used to generate a parameterized mixed state defined in Eq.(\ref{IV:eq5}). The number of parameters of Algorithm 2 is $O(Sn^2)$. Figure~\ref{Fig2} shows the schematic illustration of both adaptive algorithms.

The mixed parameterized quantum states in both algorithms are constructed as a linear combination of parameterized pure states. Without loss of generality, we denote both states as a general form, $\sigma(\boldsymbol{q},\boldsymbol{\theta})=\sum_{l=1}^{n_l}q_l\sigma_l=\sum_{l=1}^{n_l}q_l\mathcal{U}_l(\theta_l)|0^{\otimes n}\rangle\langle0^{\otimes n}|\mathcal{U}_l^{\dag}(\theta_l)$, where $\mathcal{U}_l(\boldsymbol{\theta})$ are PQCs and $|0^{\otimes n}\rangle$ is the initial state. In real quantum devices, we only require to apply each unitary $\mathcal{U}_l(\theta_l)$ on $|0^{\otimes n}\rangle$ to generate the pure state $\sigma_l=\mathcal{U}_l(\theta_l)|0^{\otimes n}\rangle\langle0^{\otimes n}|\mathcal{U}_l^{\dag}(\theta_l)$. Together with classical parameters $\{q_l\}$, we prepare the ensemble $\{q_l,\sigma_l\}$. This strategy is feasible for NISQ devices since the unitaries $\mathcal{U}_l(\theta_l)$ belong to the pools $\mathcal{P}_1$, $\mathcal{P}_2$, and thus are efficiently applicable for NISQ devices. Note that this strategy has been used to compile mixed states~\cite{ezzell2023quantum}.

\section{Numerical simulations}
\begin{figure}[ht]
    \centering
    \includegraphics[scale=0.55]{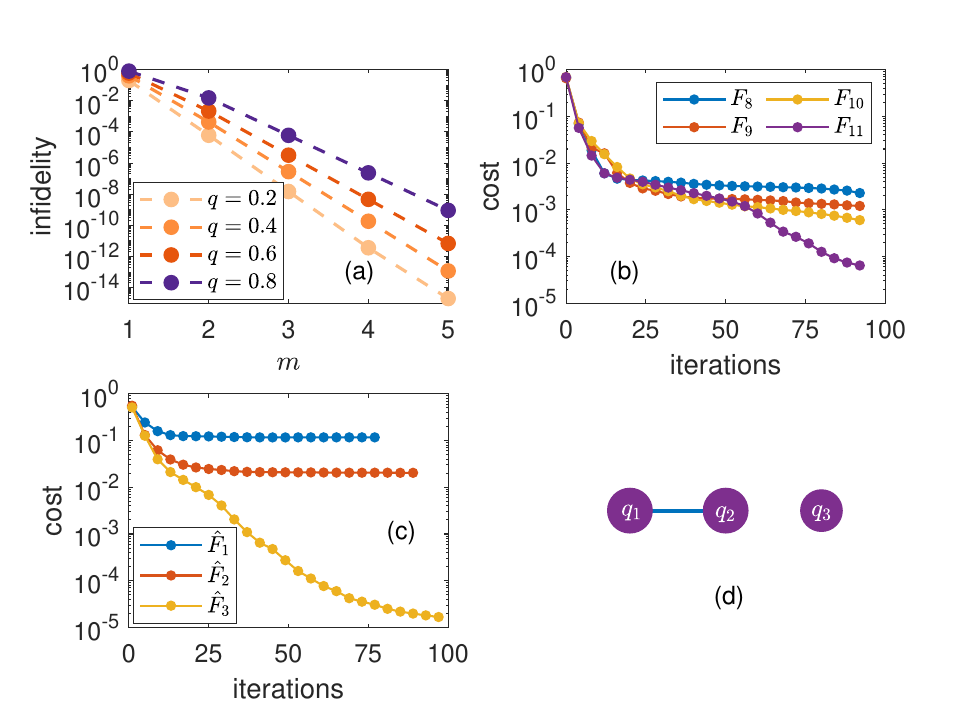}
    \caption{(a) The infidelity defined in Eq.(\ref{V:eq2}) as a function of $m$ for different noise strength $q$. (b) The training process for learning the $4$-qubit mixed state $\varrho_4(0.9)$ defined by Eq.(\ref{V:eq3}). (c) The training process for learning the $3$-qubit mixed state $\varrho_3(0.7)$ defined by Eq.(\ref{V:eq3}). (d) The graph of the first pure state in the learned decomposition, Eq.(\ref{V:eq4}).}
    \label{Fig3}
\end{figure}

We simulate the proposed algorithms in an ideal, noise-free setting via state-vector simulation. Consider a $4$-qubit mixed state $\varrho_4(q)=\varrho_3(q)\otimes|0\rangle\langle0|$, where the $3$-qubit mixed state
\begin{align}\label{V:eq3}
    \varrho_3(q)&=(1-q)|\mbox{GHZ}_3\rangle\langle\mbox{GHZ}_3|+\frac{q}{8}\mathds{1}_8.
\end{align}
The mixed state $\varrho_3(q)$ is a fully separable state if and only if $q\in[\frac{4}{5},1]$~\cite{guhne2010separability}. Therefore, the state $\varrho_4(0.9)$ is a fully separable mixed state. We employ Algorithm 1 to learn the state $\varrho_4(0.9)$ by selecting $S=8,9,10,11$ shown in Fig.~\ref{Fig3}(b). The minimization value of the cost function $F_{11}$ is lower than the threshold $10^{-4}$. Note that $11(1+2\times4)=99$ parameters are sufficient to detect that the state $\varrho_4(q)$ is a fully separable mixed state. As a comparison, the fixed parametrization uses $2^4(1+2\times4)=144$ parameters~\cite{consiglio2022variational}. Moreover, the rank of the state $\varrho_4(0.9)$ is $8$, which is not a full-rank state. Thus, our adaptive methods are more efficient for low-rank states than the fixed parametrization~\cite{consiglio2022variational}.
    
Figure~\ref{Fig3}(c) shows the optimization process of the state $\varrho_3(0.7)$ by setting $S=1,2,3$ in Algorithm 2. When $S=3$, the minimal value of the cost function $\hat{F}_3$ is lower than a threshold $10^{-4}$. Note that there are three PQC in $\mathcal{P}_2$, in which

\begin{align}
    V_1(\boldsymbol{\alpha},\boldsymbol{\theta})&=W(\boldsymbol{\alpha})\left[Q_{1,2}(\boldsymbol{\theta}_{i})\otimes\eins_{2}\right],\nonumber\\
    V_2(\boldsymbol{\alpha},\boldsymbol{\theta})&=W(\boldsymbol{\alpha})\left[\eins_{2}\otimes Q_{2,3}(\boldsymbol{\theta}_{i})\right],\nonumber\\
    V_3(\boldsymbol{\alpha},\boldsymbol{\theta})&=W(\boldsymbol{\alpha})\left[\eins_{2}\otimes Q_{1,3}(\boldsymbol{\theta}_{i})\right].
\end{align}
Hence, the state $\varrho_3(0.7)$ is either a $2$-separable or a $3$-separable state. Denote the learned decomposition as
\begin{align}\label{V:eq4}
    \sigma(\boldsymbol{\Phi}^{*})=\sum_{m=1}^{9}q_m^{*}\sigma_m^{*},
\end{align}
where the pure states $\sigma_m^{*}$ for $m=1,2,3$ are generated by the unitary $V_1$, $\sigma_m^{*}$ for $m=4,5,6$ are generated by the unitary $V_2$, and $\sigma_m^{*}$ for $m=7,8,9$ are generated by the unitary $V_3$. We focus on the first pure state $\sigma_1^{*}$ and check the entanglement of the qubit pair $(1,2)$ by using the method in Sec.~\ref{SecIII:A}. Figure~\ref{Fig3}(d) implies that the qubit pair $(1,2)$ is entangled since the purity of the reduced state by tracing the second and the third qubits is $0.6155\neq1$. As a result, the pure state $\sigma_1^{*}$ is a $2$-separable state since the corresponding graph of the state $\sigma_1^{*}$ can be divided into two subgraphs. The result indicates that $\varrho_3(0.7)$ is a $2$-separable state that agrees with the exact result.

\section{Conclusion and outlook}
In summary, we have established a link between PQCs and the separability of the state. If the target state can be learned by the PQCs in the constructed PQC pools, our approach can determine its separability by only checking the entanglement between two-qubit interaction unitary. In particular, the number of two-qubit interaction unitary required to check is at most $\lfloor n/2\rfloor$. In comparison, the direct approach needs to measure $n(n-1)/2$ qubit pairs to determine the separability of an $n$-qubit pure state. This reduction makes our approach more feasible for detecting the entanglement of pure states. Note that our approach is suitable for states that have optimal PQC from the pools. Moreover, we address the separability detection of pure states even though their preparation has the global depolarizing noise. This result provides possibilities to analyze the pure state entanglement from the obtained noisy mixed state. Finally, we have presented two adaptive algorithms to determine the separability of mixed states. Our strategies are efficient for low-rank states, such as $R(\vr)=\Omega(n)$.
    
Our findings suggest several areas for future research. First, our approach may encounter the barren plateaus~\cite{mcclean2018barren}. It would be interesting to introduce other strategies into our algorithms to improve their efficiency of our algorithms, such as defining local cost functions~\cite{ezzell2023quantum,cerezo2021cost}. Second, the application of purification-based quantum error mitigation techniques~\cite{koczor2021exponential,huggins2021virtual} could be explored in other entanglement detection methods~\cite {chen2023near,liu2023solving}. Finally, our adaptive methods are based on the probability ensemble ansatz. As suggested in~\cite{yin2022efficient}, the purification ansatz can be generalized to solve the problem of fully separable mixed states with resource requirements scaling polynomially with the number of qubits.
	
\section*{ACKNOWLEDGMENTS}
This work is supported by the National Natural Science Foundation of China (Grant Nos. 12125402, 12347152, 12405006, 12075159, and 12171044), Beijing Natural Science Foundation (Grant No. Z240007), the Innovation Program for Quantum Science and Technology (No. 2021ZD0301500), the Postdoctoral Fellowship Program of CPSF (No. GZC20230103), the China Postdoctoral Science Foundation (No. 2023M740118), the specific research fund of the Innovation Platform for Academicians of Hainan Province.

\section*{DATA AVAILABILITY}
The data that support the findings of this article are openly available~\cite{liang_code}.

\appendix
\section{The parameterized quantum circuit pool}\label{AppendixA}
In the main text, we introduce two PQC pools
\begin{align}
    \mathcal{P}_1&=\left\{W(\boldsymbol{\alpha})\right\},\\
    \mathcal{P}_2&=\left\{V_l\left(\boldsymbol{\alpha},\boldsymbol{\theta}\right)W\left(\boldsymbol{\gamma}\right)\right\}_{l=1}^{L}.
\end{align}
Here, we present more details on the PQCs.
		
For an arbitrary single-qubit unitary $U\in\mathbf{SU}(2)$, there exist real numbers $x_1$, $x_2$, $x_3$, and $x_4$ such that~\cite{nielsen2010quantum}
\begin{align}\label{AppA:eq1}
    U=e^{\iota x_1}R_z(x_2)R_y(x_3)R_z(x_4),~\iota^2=-1,
\end{align}
where rotations are about the $y$ and $z$ axes, respectively. Ignoring the global phase, we obtain a parametrization of a single-qubit unitary,
\begin{align}\label{AppA:eq2}
    U(\boldsymbol{\alpha})=R_z(\alpha_3)R_y(\alpha_2)R_z(\alpha_1).
\end{align}
If we only consider the fully separable states, the parametrization can further be reduced as $U(\boldsymbol{\alpha})=R_z(\alpha_2)R_y(\alpha_1)$. For $k$-separability we still use Eq.(\ref{AppA:eq1}) as a parametrization of a single-qubit unitary.
		
For a two-qubit system, it is well known that any $U\in\mathbf{SU}(4)$ can be written as~\cite{kraus2001optimal,zhang2003geometric,shende2004minimal}
\begin{align}\label{AppA:eq3}
    U=\left(U_1^{2}\otimes U_2^{2}\right)e^{\iota\mathcal{H}}\left(U_1^{1}\otimes U_2^{1}\right),
\end{align}
where the Hamiltonian $\mathcal{H}=\theta_{x}X\otimes X+\theta_{y}Y\otimes Y+\theta_{z}Z\otimes Z$ and each unitary $U_i^{j}\in\mathbf{SU}(2)$ that can be parameterized by Eq.(\ref{AppA:eq1}). Moreover, the unitary $e^{\iota\mathcal{H}}$ can be decomposed as (see Fig. 6 in Ref.~\cite{vatan2004optimal})
\begin{align}\label{AppA:eq4}
    e^{\iota\mathcal{H}}=\left[\mathds{1}\otimes R_z\left(\frac{\pi}{2}\right)\right]Q(\boldsymbol{\theta})\left[R_z\left(-\frac{\pi}{2}\right)\otimes\mathds{1}\right],
\end{align}
where
\begin{align}\label{AppA:eq5}
    Q(\boldsymbol{\theta})&=\mathrm{C}_{2,1}\left[ R_z\left(2\theta_z-\frac{\pi}{2}\right)\otimes R_y\left(\frac{\pi}{2}-2\theta_x\right)\right]\nonumber\\
    &\times\mathrm{C}_{1,2}\left[\mathds{1}\otimes R_y\left(2\theta_y-\frac{\pi}{2}\right)\right]\mathrm{C}_{2,1},
\end{align}
and the unitary $\mathrm{C}_{i,j}$ means the controlled-not gate where the controlled qubit is $i$ and the work qubit is $j$.
    
Hence, the two-qubit PQC pools are $\mathcal{P}_1=\{W(\boldsymbol{\alpha})\}$, $\mathcal{P}_2=\{V_1\left(\boldsymbol{\alpha}^{1},\boldsymbol{\theta}^{1}\right)W\left(\boldsymbol{\alpha}^{0}\right)\}$, where unitaries $W(\boldsymbol{\alpha})=U(\boldsymbol{\alpha}_1)\otimes U(\boldsymbol{\alpha}_2)$ and $V_1(\boldsymbol{\alpha}^{1},\boldsymbol{\theta}^{1})=W(\boldsymbol{\alpha}^{1})Q(\boldsymbol{\theta}^{1})$. Note that PQC from $\mathcal{P}_1$ can only be used to generate fully separable pure states. PQC from $\mathcal{P}_2$ can generate entangled two-qubit states.

Now, we show the construction of $\{V_l\}_{l=1}^{L}$ for an $n$-qubit system. The basic idea is to determine which two qubits are connected. Thus, we only need to arrange the qubits. We first assume $n$ is even.

(1) $n=2$, the connected qubit pair is $(1,2)$. Then only one unitary $V_1$ is avaiable.

(2) $n=4$, all possible qubit pairs can be arranged in three rows and three columns,
\begin{align}
    &\textcolor{blue}{(1,2)},\textcolor{blue}{(1,3)},\textcolor{blue}{(1,4)};\nonumber\\
    &\textcolor{violet}{(2,3)},\textcolor{blue}{(2,4)};\nonumber\\
    &\textcolor{blue}{(3,4)}.
\end{align}
The first column can be split into two unitaries since $(1,2)$ and $(3,4)$ can be manipulated simultaneously. The second column can be treated as a single unitary since $(1,2)$ and $(3,4)$ can also be manipulated simultaneously. The third column is a single unitary. For each column, the qubit pairs with the same color can be manipulated simultaneously. Thus, there are $2\times(\frac{4}{2}-1)+\frac{4}{2}=4$ unitaries. Note that $(1,4)$ and $(2,3)$ can be manipulated simultaneously. As a result, the total number of unitaries is $L=2\times(\frac{4}{2}-1)+\frac{4}{2}-1=3$.

(3) $n=6$, all possible qubit pairs can be arranged in five rows and five columns,
\begin{align}
    &\textcolor{blue}{(1,2)},\textcolor{blue}{(1,3)},\textcolor{blue}{(1,4)},\textcolor{blue}{(1,5)},\textcolor{blue}{(1,6)};\nonumber\\
    &\textcolor{violet}{(2,3)},\textcolor{blue}{(2,4)},\textcolor{blue}{(2,5)},\textcolor{blue}{(2,6)};\nonumber\\
    &\textcolor{blue}{(3,4)},\textcolor{violet}{(3,5)},\textcolor{blue}{(3,6)};\nonumber\\
    &\textcolor{violet}{(4,5)},\textcolor{violet}{(4,6)};\nonumber\\
    &\textcolor{blue}{(5,6)}.
\end{align}
The total number of unitaries is $L=2\times(\frac{6}{2}-1)+\frac{6}{2}-1=6$.

(4) $n=8$, all possible qubit pairs can be arranged in seven rows and seven columns,
\begin{align}
    &\textcolor{blue}{(1,2)},\textcolor{blue}{(1,3)},\textcolor{blue}{(1,4)},\textcolor{blue}{(1,5)},\textcolor{blue}{(1,6)},\textcolor{blue}{(1,7)},\textcolor{blue}{(1,8)};\nonumber\\
    &\textcolor{violet}{(2,3)},\textcolor{blue}{(2,4)},\textcolor{blue}{(2,5)},\textcolor{blue}{(2,6)},\textcolor{blue}{(2,7)},\textcolor{blue}{(2,8)};\nonumber\\
    &\textcolor{blue}{(3,4)},\textcolor{violet}{(3,5)},\textcolor{blue}{(3,6)},\textcolor{blue}{(3,7)},\textcolor{blue}{(3,8)};\nonumber\\
    &\textcolor{violet}{(4,5)},\textcolor{violet}{(4,6)},\textcolor{violet}{(4,7)},\textcolor{blue}{(4,8)};\nonumber\\
    &\textcolor{blue}{(5,6)},\textcolor{blue}{(5,7)},\textcolor{violet}{(5,8)};\nonumber\\
    &\textcolor{violet}{(6,7)},\textcolor{blue}{(6,8)};\nonumber\\
    &\textcolor{blue}{(7,8)}.
\end{align}
The total number of unitaries is $L=2\times(\frac{8}{2}-1)+\frac{8}{2}-1=9$.

\begin{figure}[ht]
    \centering
    \includegraphics[scale=1]{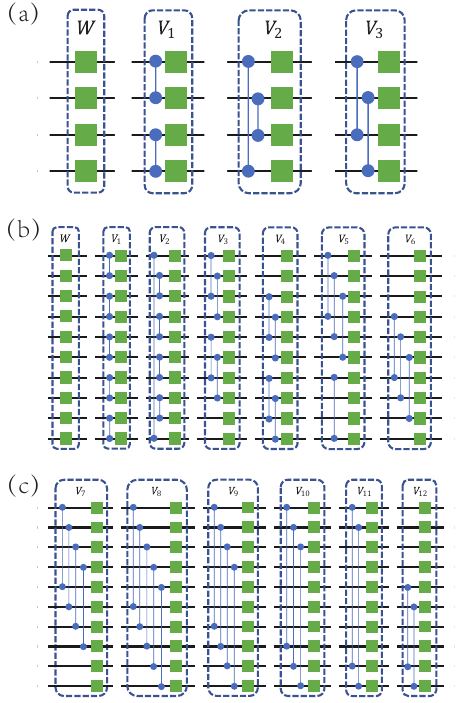}
    \caption{The PQC pools for (a) $4$- and (b,c) $10$-qubit systems. The dot-line-dot denotes the two-qubit unitary $Q$ defined in Eq.(\ref{II:eq1}).}
    \label{SFig1}
\end{figure}

In summary, the total number of unitaries of an $n$-qubit system is
\begin{align}
    L=2\times(\frac{n}{2}-1)+\frac{n}{2}-1=\frac{3n}{2}-1.
\end{align}
Figure~\ref{SFig1} shows examples for $4$- and $10$-qubit systems. The first two unitaries $V_1$ and $V_2$ follow the definition of Eqs.(\ref{II:eq4}) and (\ref{II:eq5}) such as
\begin{align}
    V_1(\boldsymbol{\alpha},\boldsymbol{\theta})
    \!&\!=\!W(\boldsymbol{\alpha})\left[\bigotimes_{i=1}^{n-1}Q_{i,i+1}(\boldsymbol{\theta}_{i})\right],\\
    V_2(\boldsymbol{\alpha},\boldsymbol{\theta})
    \!&\!=\!W(\boldsymbol{\alpha})\left[Q_{1,n}(\boldsymbol{\theta}_{1})\otimes\left(\bigotimes_{j=2}^{n-2}Q_{j,j+1}(\boldsymbol{\theta}_{j})\right)\right],
\end{align}
where $i$ runs over odd numbers and $j$ runs over even numbers.

For odd $n$, we consider $n=9$. All possible qubit pairs can be arranged in eight rows and eight columns:
\begin{align}
    &\textcolor{blue}{(1,2)},\textcolor{blue}{(1,3)},\textcolor{blue}{(1,4)},\textcolor{blue}{(1,5)},\textcolor{blue}{(1,6)},\textcolor{blue}{(1,7)},\textcolor{blue}{(1,8)},\textcolor{blue}{(1,9)};\nonumber\\
    &\textcolor{violet}{(2,3)},\textcolor{blue}{(2,4)},\textcolor{blue}{(2,5)},\textcolor{blue}{(2,6)},\textcolor{blue}{(2,7)},\textcolor{blue}{(2,8)},\textcolor{blue}{(2,9)};\nonumber\\
    &\textcolor{blue}{(3,4)},\textcolor{violet}{(3,5)},\textcolor{blue}{(3,6)},\textcolor{blue}{(3,7)},\textcolor{blue}{(3,8)},\textcolor{blue}{(3,9)};\nonumber\\
    &\textcolor{violet}{(4,5)},\textcolor{violet}{(4,6)},\textcolor{violet}{(4,7)},\textcolor{blue}{(4,8)},\textcolor{blue}{(4,9)};\nonumber\\
    &\textcolor{blue}{(5,6)},\textcolor{blue}{(5,7)},\textcolor{violet}{(5,8)},\textcolor{violet}{(5,9)};\nonumber\\
    &\textcolor{violet}{(6,7)},\textcolor{blue}{(6,8)},\textcolor{violet}{(6,9)};\nonumber\\
    &\textcolor{blue}{(7,8)},\textcolor{violet}{(7,9)};\nonumber\\
    &\textcolor{violet}{(8,9)}.
\end{align}
The total number of unitaries is $L=2\times\frac{9-1}{2}+\frac{9-1}{2}=12$. In summary, the total number of unitaries of an $n$-qubit system is
\begin{align}
    L=3\times\frac{n-1}{2}.
\end{align}

Finally, we remark that not all pure states can be learned from our approach. Based on the construction of PQCs, all $(n-1)$ separable states and $2$-producible states with nearest-neighbor connections can be verified.

\bibliography{refs}

\end{document}